%% file: main.tex
\documentclass[conference]{IEEEtran}
\IEEEoverridecommandlockouts

\usepackage{cite}
\usepackage{amsmath,amssymb,amsfonts,amsbsy}
\usepackage{algorithmic}
\usepackage{graphicx}
\usepackage{textcomp}
\usepackage{xcolor}
\usepackage[mode=tex]{standalone} 
\usepackage{subfig}
\usepackage{balance}

\def\BibTeX{{\rm B\kern-.05em{\sc i\kern-.025em b}\kern-.08em
    T\kern-.1667em\lower.7ex\hbox{E}\kern-.125emX}}

\input{commands.tex}

\begin{document}

\title{
	DFT-p-FDMA Based Chirp Transmission in CP-OFDM for Unified ISAC Waveform Design
}

\author{
	\IEEEauthorblockN {
		Fabrizio Carpi, 
		Joonyoung Cho, 
		Kyeong Jin Kim, 
		Charlie Jianzhong Zhang
	}
	\IEEEauthorblockA{
		Standards and Mobility Innovation, Samsung Research America, USA \\
		Email: \{f.carpi, joonyoung.cho, kj.kim1, jianzhong.z\}@samsung.com
	}
}

\maketitle

\begin{abstract}
We propose an integrated sensing and communications (ISAC) framework that supports chirp signal transmission in CP-OFDM-based multiple access communication systems, enabling efficient coexistence of communication and sensing capabilities.
Our framework employs the discrete Fourier transform \emph{phase rotated and permuted} frequency division multiple access (DFT-p-FDMA) waveform~\cite{DFTpFDMA-2025} to transmit chirp signals using a portion of the frequency resources, while ensuring interference-free concurrent CP-OFDM data transmissions on other bands.
We analyze the effective channel behavior under the DFT-p-FDMA waveform, characterizing how delays and Doppler shifts impact radar target echoes.
We also show how processing multiple received symbols improves Doppler resolution in practical scenarios.
Our framework allows flexible adjustment of range-Doppler resolution through optimized time-frequency resource allocation, offering a versatile solution for ISAC applications.
Simulation results validate the framework’s performance in delay and Doppler estimation, highlighting its potential to support ISAC in next-generation wireless networks.
\end{abstract}

\vspace{-0.55em}

\section{Introduction}

	Beyond their core role in data transmission, emerging wireless networks are increasingly expected to incorporate sensing capabilities while maintaining backward compatibility with the existing infrastructure.
	In this context, 6G is anticipated to support the integrated sensing and communications (ISAC) paradigm, which leverages the use of cellular networks to perform radar-like tasks such as localization and environmental mapping.
	By capitalizing on the widespread coverage and high bandwidth of cellular systems, ISAC facilitates comprehensive sensing, eliminating the requirement for dedicated radar equipment or additional spectrum allocation.
	This integration effectively turns the communication network into a sensor, providing critical data for sensing applications.
	
	Modern radar systems often utilize linear frequency modulation, commonly known as chirp waveform, due to its capabilities in ranging applications~\cite{RadarBook}.
	For instance, automotive radars extensively employ chirp-based solutions to achieve high-resolution detection and ranging that enable precise localization of vehicles and obstacles in real-time~\cite{Pator-AutomotiveRadars-2017}.
	Beyond automotive applications, chirp-based radar has also been investigated for unmanned aerial vehicle (UAV) detection, as it effectively resolves small, fast-moving targets in cluttered environments~\cite{Oh-DroneDet-2021}.
	This adaptability highlights the versatility of chirp-based radar, from ground-based vehicular systems to aerial surveillance applications.
	One key benefit of chirp-based radar is the decoupling between pulse duration and range resolution, enabling transmission of high-energy long-duration pulses that achieve fine range resolution.
	Additionally, the chirp signal's constant envelope characteristic results in a minimal peak-to-average power ratio (PAPR) that enhances the power amplifier efficiency.
	
	Chirp signals can be generated using the discrete affine Fourier transform (DAFT) waveform~\cite{DAFT-2005}, which employs multiple orthogonal chirps.
	Building upon the DAFT framework, the affine frequency division multiplexing (AFDM) waveform~\cite{AFDM-2023} optimizes the DAFT parameters to achieve delay-Doppler diversity, enabling communications in doubly selective channels.
	Another chirp waveform scheme is the discrete Fourier transform \emph{phase rotated and permuted} frequency domain multiple access (DFT-p-FDMA)~\cite{DFTpFDMA-2025}, which provides a framework to multiplex a chirp waveform and cyclic prefix orthogonal frequency division multiplexing (CP-OFDM).
	DFT-p-FDMA enables the generation of a chirp signal by incorporating a pre-processing that includes a DFT block and a phase-rotated permutation prior to the CP-OFDM IDFT.
	Similar to AFDM, DFT-p-FDMA also leverages signals in the DAFT domain, facilitating the delay-Doppler processing that is highly suitable for ISAC applications.

	Recently, several works have studied the AFDM-based ISAC systems.
	Pilot design and receiver aspects are discussed in~\cite{Bemani-ISAC-AFDM-2024}.
	In~\cite{Ni-ISAC-AFDM-2025}, communications and sensing metrics are analyzed and an estimation method for delay and Doppler is presented. 
	In~\cite{Ranasinghe-2025}, communications and sensing algorithms are analyzed comparing AFDM, OFDM, and orthogonal time frequency space (OTFS) modulation.

	In this paper, we propose an ISAC framework utilizing the DFT-p-FDMA waveform within CP-OFDM-based systems.
	Our contributions are summarized as follows:
	\begin{itemize}
		\item We derive closed-form expressions for the effective channel under the DFT-p-FDMA waveform, rigorously characterizing the impact of delay and Doppler shifts on radar target echoes (Section~\ref{sec:system-model}).
		The theoretical analysis is supported by illustrative examples and visualizations, offering deeper insights into the system behavior.
		
		\item We analyze a receiver scheme that leverages multiple symbols to enhance the Doppler resolution, thereby improving target parameter estimation performance (Section~\ref{sec:Rx-processing}).
		
		\item Through extensive simulations, we validate the received signal characteristics and evaluate the performance of delay and Doppler estimation under the point scatterer model (Section~\ref{sec:results}).
	\end{itemize}

\section{System Model}
\label{sec:system-model}

	We consider the DFT-p-FDMA waveform, which enables the generation of a chirp signal by incorporating a pre-processing prior to the CP-OFDM framework of 4G/5G systems~\cite{DFTpFDMA-2025}.
	At the transmitter side, the pre-processing includes a DFT block and a phase-rotated permutation, and these operations are reversed with the post-processing at the receiver.
	DFT-p-FDMA supports in-band ISAC processing without causing interference to other sub-bands within the CP-OFDM framework.

	\begin{figure*}
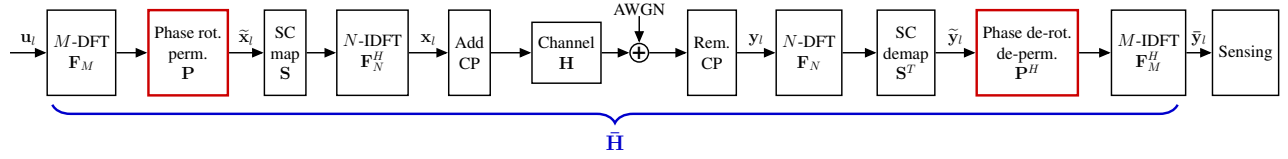

		\centering
		\includestandalone[width=1.9\columnwidth]{Block_diagram_DFTpFDMA}
		\vspace{-0.5em}
		\caption{System model for DFT-p-FDMA.}
		\label{fig:system-model}
		\vspace{-1.5em}
	\end{figure*}

	The system model is shown in Fig.~\ref{fig:system-model}.
	The input sequence $\vect{u}_l=\left[u_l[0],\dots,u_l[M-1]\right]^T$, whose elements $u_l[m]\in\mathbb{C}$, is defined across the DAFT-domain indices $m=0,\dots,M-1$ and the OFDM symbol indices $l=0,\dots,L-1$. 
	Here, $M$ represents the number of OFDM subcarriers and $L$ denotes the total number of OFDM symbols within one transmission time interval (TTI).

	The input sequence is pre-processed with an $M$-point DFT and the phase rotated permutation matrix as
	\begin{equation}
		\vect{\xtilde}_l = \mat{P} \mat{F}_M \vect{u}_l,
		\label{eq:x-tilde}
	\end{equation}
	where $\vect{F}_M\in\mathbb{C}^{M\times M}$ is the $M$-point DFT matrix with elements $\vect{F}_M[k,i]=e^{-j2\pi ki/M}/\sqrt{M}$, and $\mat{P}\in\mathbb{C}^{M\times M}$ is the phase rotated permutation matrix.

	The phase rotated permutation matrix $\mat{P}$ enables the spreading of the input sequence over the time-frequency resource~\cite{DFTpFDMA-2025}.
	It is defined as
	\begin{equation}
		\mat{P} = \mat{F}_M \mat{\Lambda}_c^H \mat{F}_M^H \mat{\Lambda}_c^H \mat{F}_M^H,
		\label{eq:Perm-mat}
	\end{equation}
	where $\mat{\Lambda}_c=\text{diag}\left(e^{j2\pi c m^2/2M}, m=0,\dots,M-1 \right)$ is a diagonal matrix, and $c\in\mathbb{Z}_{>0}$ is a positive integer that represents the chirp rate~\cite{DFTpFDMA-2025}.
	$\mat{P}$ is a unitary matrix such that $\mat{P}\mat{P}^H=\mat{P}^H \mat{P}=\mat{I}$, and $\mat{P}$ has only one non-zero element (unit norm complex exponential) for each column/row. 
	Additional details about $\mat{P}$ are discussed in~\cite{DFTpFDMA-2025}.
	
	The phase rotated permutation matrix $\mat{P}$ has a sparse structure, where the permutation $\pi_{\mat{P}}$ induced by $\mat{P}$ determines the locations of the non-zero elements, such that $\pi_{\mat{P}}(m)=\arg\max_l \mat{P}[m,l]$.
	The elements of $\mat{P}$ can be defined as
	\begin{equation}
		\mat{P}[m,k] = e^{j\theta(m)} \delta_{k,\pi_{\mat{P}}(m)},
		\label{eq:P-element}
	\end{equation}
	where $\theta(m)$ is the phase of the non-zero element in the $m$-th row of $\mat{P}$, 
	and $\delta_{a,b}$ is the Kronecker delta.

	After the pre-processing, the subcarrier mapping and the $N$-point IDFT are performed as
	\begin{equation}
		\vect{x}_l = \mat{F}_N^H \mat{S} \vect{\xtilde}_l,
		\label{eq:x}
	\end{equation}
	where $\mat{S}\in\{0,1\}^{N\times M}$ denotes the subcarrier mapping matrix that contains only one non-zero element for each column, and $\vect{F}_N^H\in\mathbb{C}^{N\times N}$ denotes the $N$-point IDFT matrix.
	A CP of length $\Ncp$ is added to the front of $\vect{x}_l$.  
	When the parameter $c$ yields valid permutation matrices, the CP is equivalent to a \emph{chirp-periodic} prefix (CPP) that ensures periodicity to combat multipath propagation~\cite{DFTpFDMA-2025,AFDM-2023}.
	The equivalence between CP and CPP also prevents DFT-p-FDMA signals from interfering with concurrent CP-OFDM data transmissions. 
	
	At the receiver side, the reverse operations are performed as shown in Fig.~\ref{fig:system-model}.
	The received signal for the $l$-th OFDM symbol, after CP removal, is 
	\begin{equation}
		\vect{y}_l = \mat{H}_l \vect{x}_l + \vect{z}_l,
	\end{equation}
	where $\vect{z}_l\sim\mathcal{CN}(0,2\sigma_{\text{AWGN}}^2\mat{I})$ is the complex additive white Gaussian noise (AWGN) with zero mean and covariance matrix $2\sigma_{\text{AWGN}}^2\mat{I}$.
	The $N$-point DFT and the subcarrier demapping are performed resulting in 
	\begin{equation}
		\vect{\ytilde}_l = \mat{S}^T \mat{F}_N \vect{y}_l.
		\label{eq:y-tilde}
	\end{equation}
	
	Then, a post-processing comprising the phase de-rotation de-permutation and an $M$-point IDFT is performed, resulting in the post-processing signal
	\begin{equation}
		\vect{\ybar}_l = \mat{F}_M^H \mat{P}^H \vect{\ytilde}_l.
		\label{eq:y-bar0}
	\end{equation}

	The effective DFT-p-FDMA channel is denoted with $\mat{\Hbar}_l\in\mathbb{C}^{M\times M}$ (highlighted in blue in Fig.~\ref{fig:system-model}), and it can be expressed as
	\begin{equation}
		\mat{\Hbar}_l = \mat{F}_M^H \mat{P}^H \mat{S}^T \mat{F}_N\; \mat{H}_l\; \mat{F}_N^H \mat{S} \mat{P} \mat{F}_M,
		\label{eq:H-bar}
	\end{equation}
	where $\mat{H}_l$ denotes the underlying channel for the $l$-th OFDM symbol, and the other matrices represent the Tx/Rx operations described above.
	The post-processing signal  $\vect{\ybar}_l=\left[\ybar_l[0],\dots,\ybar_l[M-1]\right]^T$ can be also expressed as 
	\begin{equation}
		\vect{\ybar}_l = \mat{\Hbar}_l \vect{u}_l + \vect{\bar{z}},
		\label{eq:y-bar}
	\end{equation}
	where $\vect{\bar{z}}\sim\mathcal{CN}(0,2\sigma_{\text{AWGN}}^2\mat{I})$ is the AWGN after post-processing, which has the same statistics as $\vect{z}$. 
	The $m$-th element $\ybar_l[m]$ of $\vect{\ybar}_l$ represents the post-processing signal component for the DAFT-domain index $m\in\{0,\dots,M-1\}$ and OFDM symbol $l\in\{0,\dots,L-1\}$.

\subsection{Channel Model}
\label{sec:channel-model}
	Consider $P$ targets denoted by $p=0,…,P-1$, each resulting in a received echo with sample delay $d_p$ and normalized Doppler $\nu_p$, for the $l=0,\dots,L-1$ OFDM symbol. 
	For a given subcarrier spacing $\Delta f$ and carrier frequency $f_c$ in a monostatic radar scenario, the range of the $p$-th target is $R_p=\frac{c_0 d_p}{2N\Delta f}$ ($c_0$ is the speed of light), its Doppler frequency is $f_{D,p} = \nu_p\Delta f$, and its relative velocity is $v_{\text{rel},p}=(c_0 f_D)/(2f_c)$.
	To simplify the mathematical representation, assume that the phase is continuous between OFDM symbols $l=0,…,L-1$. 
	The channel can be expressed as 
	\begin{equation}
		\mat{H}_l= \sum_{p=0}^{P-1} h_p \mat{\Delta}_p \Pi^{d_p} = \sum_{p=0}^{P-1} \gamma(d_p,\nu_p,l) \mat{D}(\nu_p) \mat{\Pi}^{d_p},
		\label{eq:H}
	\end{equation}	
	where $h_p\in\mathbb{C}$ is the channel gain;
	$\mat{\Delta}_p$ is a diagonal matrix 
    \begin{align}\begin{split}
    	\mat{\Delta}_p=\text{diag}\left(e^{ j2\pi\nu_p \left[\frac{n+\Ncp-d_p+1}{N} + l\left(1+\frac{\Ncp}{N}\right) \right] },n=0,\dots,N-1 \right)
    	    	\end{split}
    \end{align}
    that captures the contributions of the delay-Doppler pair $(d_p,\nu_p)$ and the CP to the phase rotation;
    $\mat{\Pi}^{d_p}$ is a cyclic-shift matrix such that is shifts the diagonal matrix by $d_p$ columns.
    The second part of~\eqref{eq:H} is obtained by mathematical manipulations, 
    where 
    \begin{equation}
    	\mat{D}(\nu_p)=\text{diag}\left(e^{j2\pi \nu_p n/N}, n=0,\dots,N-1 \right)
    \end{equation}
    represents the effect of the Doppler on each sample within an OFDM symbol, and  
    \begin{equation}
    	\gamma(d_p,\nu_p,l) = h_p\; e^{j2\pi\nu_p\left[\frac{\Ncp-d_p+1}{N} + l \left(1+\frac{\Ncp}{N}\right) \right]}
    	\label{eq:spreading-func}
    \end{equation}
    denotes the spreading function for a given delay-Doppler pair $(d_p,\nu_p)$ and OFDM symbol $l$.

\subsection{DFT-p-FDMA Effective Channel}
\label{sec:channel-DFTpFDMA}
	Given~\eqref{eq:H-bar} and~\eqref{eq:H}, we now provide a closed form expression for the effective channel that provides intuitions about the effects of delay and Doppler.
	Assume $N=M$ for the sake of simplicity in this analysis. 
	In this case $\mat{S}=\mat{S}^T=\mat{I}$.
	The effective channel $\mat{\Hbar}_l$, defined in~\eqref{eq:H-bar}, can be expressed as
	\begin{align}\begin{split}
		\mat{\Hbar}_l[n,i] = \frac{1}{M} \sum_{k,q} e^{j2\pi \frac{nk-iq}{M}} e^{j\left[ \theta\left(\pi_{\mat{P}}(q)\right) - \theta\left(\pi_{\mat{P}}(k)\right) \right]} \\
		\cdot\sum_{p=0}^{P-1} \gamma(d_p,\nu_p,l) e^{-j2\pi \: \pi_{\mat{P}}(q) \frac{d_p}{N}} D_N\left(\nu_p+\pi_{\mat{P}}(q)-\pi_{\mat{P}}(k)\right)
				\end{split}
		\label{eq:H-bar-closedForm}
	\end{align}
	where the kernel 
	\begin{equation}
		D_N(x) = \frac{1}{N} \sum_{n=0}^{N-1} e^{j2\pi x\frac{n}{N}} 
		= \frac{1}{N} e^{j\pi x \frac{N-1}{N}} \frac{\sin(\pi x)}{\sin(\pi x/N)}
		\label{eq:kernel-D_N}
	\end{equation}
	captures the Doppler-induced frequency spreading. 
	The derivation of~\eqref{eq:H-bar-closedForm} is provided in Appendix~\ref{sec:derivations}
	
	Assume that the input sequence is a constant pulse in DAFT domain, i.e., $\vect{u}_l=\vect{u}=[\sqrt{M},0,\dots,0]$, which corresponds to the transmission of a single chirp.
	From~\eqref{eq:y-bar}, the noiseless post-processing signal for the $l$-th OFDM symbol is $\vect{\ybar}_l=\mat{\Hbar}_l \vect{u}$, where each $m=0,\dots,M-1$ element can be expressed as
	\begin{equation}
		\ybar_l[m] 
		=\sum_{p=0}^{P-1} A(d_p,\nu_p,m) e^{j2\pi \nu_p' l},
		\label{eq:y-bar-ml-closedForm}
	\end{equation}
	where $\nu_p'=\nu_p \left(1+\frac{\Ncp}{N}\right)$, and the contribution of the delay-Doppler pairs $(d_p,\nu_p)$ to DAFT-domain index $m$ is 
	\begin{align}\begin{split}
		A(d_p,\nu_p,m) = \frac{1}{M} \sum_{k,q} e^{j2\pi m \frac{k}{M}} e^{j\left[ \theta\left(\pi_{\mat{P}}(q)\right) - \theta\left(\pi_{\mat{P}}(k)\right) \right]}\\
		\cdot \sum_{p=0}^{P-1} h_p e^{j2\pi \nu_p \frac{\Ncp-d_p}{N}} e^{-j2\pi \pi_{\mat{P}}(q)\frac{d_p}{N}} D_N(\nu_p+\pi_{\mat{P}}(q)-\pi_{\mat{P}}(k)).
				\end{split}
	\end{align}

\section{Receiver Processing}
\label{sec:Rx-processing}

	For one TTI, the post-processing signals $\vect{\ybar}_l$ consist of $m=0,\dots,M-1$ DAFT-domain samples for each of the $l=0,\dots,L-1$ OFDM symbols.
	Each DAFT-domain sample index corresponds to delay-Doppler pairs.
	Estimating Doppler from a single OFDM symbol is challenging, as it relies on the profile of the post-processing signal $\vect{\ybar}$ in the DAFT domain.
	This difficulty intensifies with multiple targets, as their echoes overlap in $\vect{\ybar}$.
	In this section we show how multiple OFDM symbols can be processed to address this challenge by enhancing the Doppler resolution. 
	
	The $L$ post-processing signals $\vect{\ybar}_l$ within one TTI are collected to form a matrix $\mat{\Ybar}\in\mathbb{C}^{M\times L}$ as
	\begin{equation}
		\mat{\Ybar} = \left[\vect{\ybar}_0,\dots,\vect{\ybar}_{L-1}\right],
	\end{equation}
	where its columns are the post-processing signals $\vect{\ybar}_l$ for each OFDM symbol $l$. 
	The matrix $\mat{\Ybar}$ captures a 2D snapshot of the post-processing signals over the DAFT domain ($M$ rows) and the time domain ($L$ columns).
	
	As in traditional radar systems~\cite{RadarBook,Sturm-WFradar-2011}, the row-wise DFT operation can be used to track the phase history of the term $e^{j2\pi\nu_p'l}$ in~\eqref{eq:y-bar-ml-closedForm} over the $L$ OFDM symbols for a given DAFT-domain index $m$.
	By applying an $L$-point DFT to the $m$-th row of $\mat{\Ybar}$, the vector $\vect{w}_m=\left[w_m[0],\dots,w_m[L-1]\right]^T$ is obtained such that for $k=0,\dots,L-1$,
	\begin{equation}
		w_m[k] = \sum_{p=0}^{P-1} \frac{A(d_p,\nu_p,m)}{\sqrt{L}} \sum_{l=0}^{L-1} e^{j2\pi l \left(\nu_p' - \frac{k}{L}\right)}.
	\end{equation}
	The matrix $\mat{W}\in\mathbb{C}^{M\times L}$ containing the rows $\vect{w}_m^T$ for $m=0,\dots,M-1$ is
	\begin{align}
		\begin{split}
			\mat{W} &= \begin{bmatrix}
				\vect{w}_0^T\\
				\vdots\\
				\vect{w}_{L-1}^T
			\end{bmatrix}.
		\end{split}
	\end{align}
	Here, $\mat{W}$ represents the 2D map that captures the DAFT domain and the Doppler domain, similar to the 2D Range-Doppler map that can be obtained in traditional radar systems~\cite{RadarBook}.
	For a single target scenario, finding the indices $(\hat{m},\hat{k})$ for which $|\mat{W}|$ is maximized corresponds to the maximum likelihood estimate of the (quantized) delay-Doppler pair~\cite{Braun-MLradarOFDM-2010}.

	Each of the $k=0,\dots,L-1$ columns of $\mat{W}$ represents a normalized Doppler bin, where the Doppler resolution is 
	\begin{equation}
		\Delta\nu=\frac{1}{L\left(1+\frac{\Ncp}{N}\right)}.
		\label{eq:Doppler-Res}
	\end{equation}
	The $k$-th normalized Doppler bin corresponds to a normalized Doppler value $\nu_p=a\cdot\Delta\nu$, which is an integer multiple of $\Delta\nu$. 
	Here, $a\in\mathbb{Z}$ is the integer multiple, and the corresponding bin index is given by $k=a\mod L$, where $\mod\star$ denotes the modulo operation.
	For example, Doppler values $\nu_p=[0,L,2L]\cdot\Delta\nu$ correspond to the Doppler bin $k=0=[0,L,2L]\mod L$ since $a=[0,L,2L]$ by the definition of $\nu_p$ in this example.
	On the other hand, Doppler values that are fractional multiples of the resolution, i.e., $\nu_p=b\cdot\Delta\nu$ where $b\in\mathbb{R}$, will manifest as a \emph{leakage} between the Doppler bins.

	The factor $\alpha_T=\left(1+\frac{\Ncp}{N}\right)$ accounts for the interval between consecutive OFDM symbols, incorporating the CP duration.
	In case non-consecutive OFDM symbols are used, the factor $\alpha_T$ needs to be adjusted to capture the interval between the symbols. 
	For example, if every other OFDM symbol is used, then $\alpha_T=2\left(1+\frac{\Ncp}{N}\right)$.

	In the remainder of this section, we present examples to provide intuitions about the structure of the post-processing signals $\vect{\ybar}$ and the 2D map $\mat{W}$. 
	In order to isolate the impact of the effective channel with DFT-p-FDMA, i.e., $\mat{\Hbar}$ defined in~\eqref{eq:H-bar} and~\eqref{eq:H-bar-closedForm}, we consider a noiseless scenario $(\sigma_\text{AWGN}^2=0)$ with $M=N=120$, $c=11$, $\Ncp=8$, $h_p=1$.
	
	\begin{figure}
		\includegraphics[width=\columnwidth]{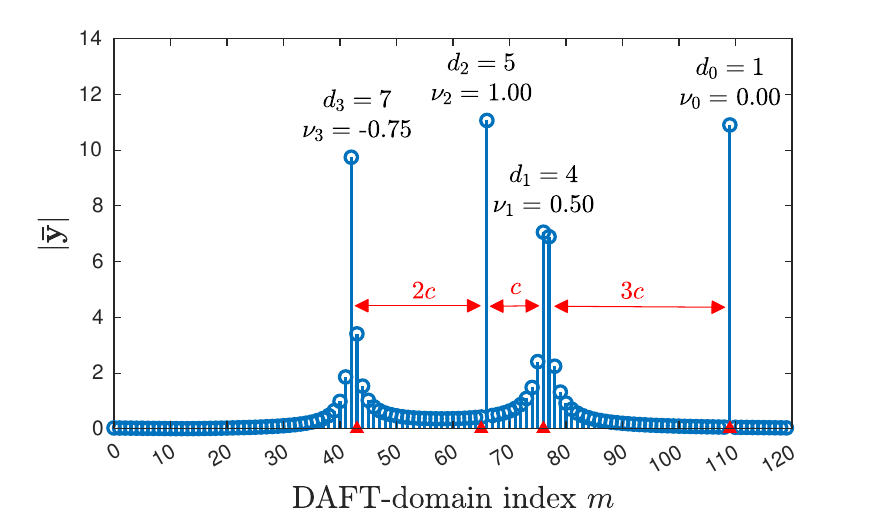}
		\caption{
			Visualization of $|\vect{\ybar}|$ for a noiseless scenario of four echoes with $d_p=[1,4,5,7]$ and $\nu_p=[0,0.5,1,-0.75]$.
			The red triangular markers on the horizontal axis denote the indices $m=-d_p\cdot c\mod M$.
		}
		\label{fig:ybar-example}
	\end{figure}

	Fig.~\ref{fig:ybar-example} shows an example of a single post-processing signal $\vect{\ybar}_0$~\eqref{eq:y-bar-ml-closedForm} in a scenario with four echoes characterized by delays $d_p=[1,4,5,7]$ and Doppler values $\nu_p=[0,0.5,1,-0.75]$.
	The red triangular markers on the horizontal axis denotes the indices $m=-d_p\cdot c\mod M$, which corresponds to echoes' locations when the Doppler is zero, where $c$ is the chirp rate parameter defined in Section~\ref{sec:system-model}. 
	This implies that pure delays (no-Doppler) are spaced by $c$.
	The effects of Doppler manifest as shift or leakage of the echoes' peaks according to the Doppler value $\nu_p$.
	For example, the echo $p=2$ is shifted to the right by one sample (with respect to the no-Doppler index denoted with the red triangle) since the Doppler $\nu_2=1$ is integer.
	For the echo $p=1$, the fractional Doppler $\nu_1=0.5$ manifests as a leakage over the DAFT-domain indices with two adjacent peaks of similar magnitudes.
	For the echo $p=3$, the fractional Doppler $\nu_3=-0.75$ manifests as a leakage too, but with a larger peak to the left of the corresponding no-Doppler index. 
	Each echo is separated by approximately multiples of $c$ depending on their delay-Doppler values $(d_p,\nu_p)$.
	
	\begin{figure}
		\centering
		\subfloat[$\nu=0.87$.]{
			\includegraphics[width=0.49\columnwidth]{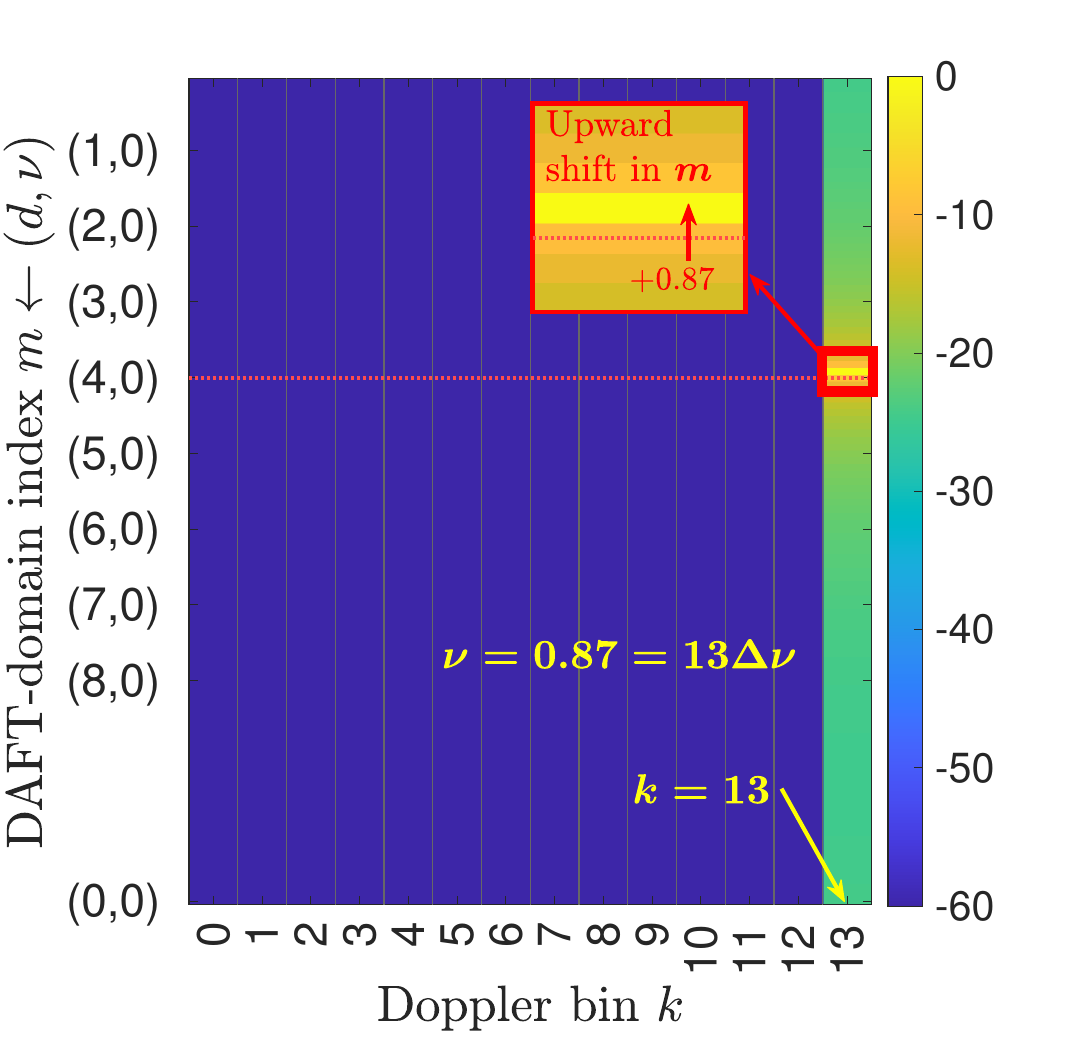}
			\label{fig:2Dmap-nu087}
		}
		\subfloat[$\nu=-0.94$.]{
			\includegraphics[width=0.49\columnwidth]{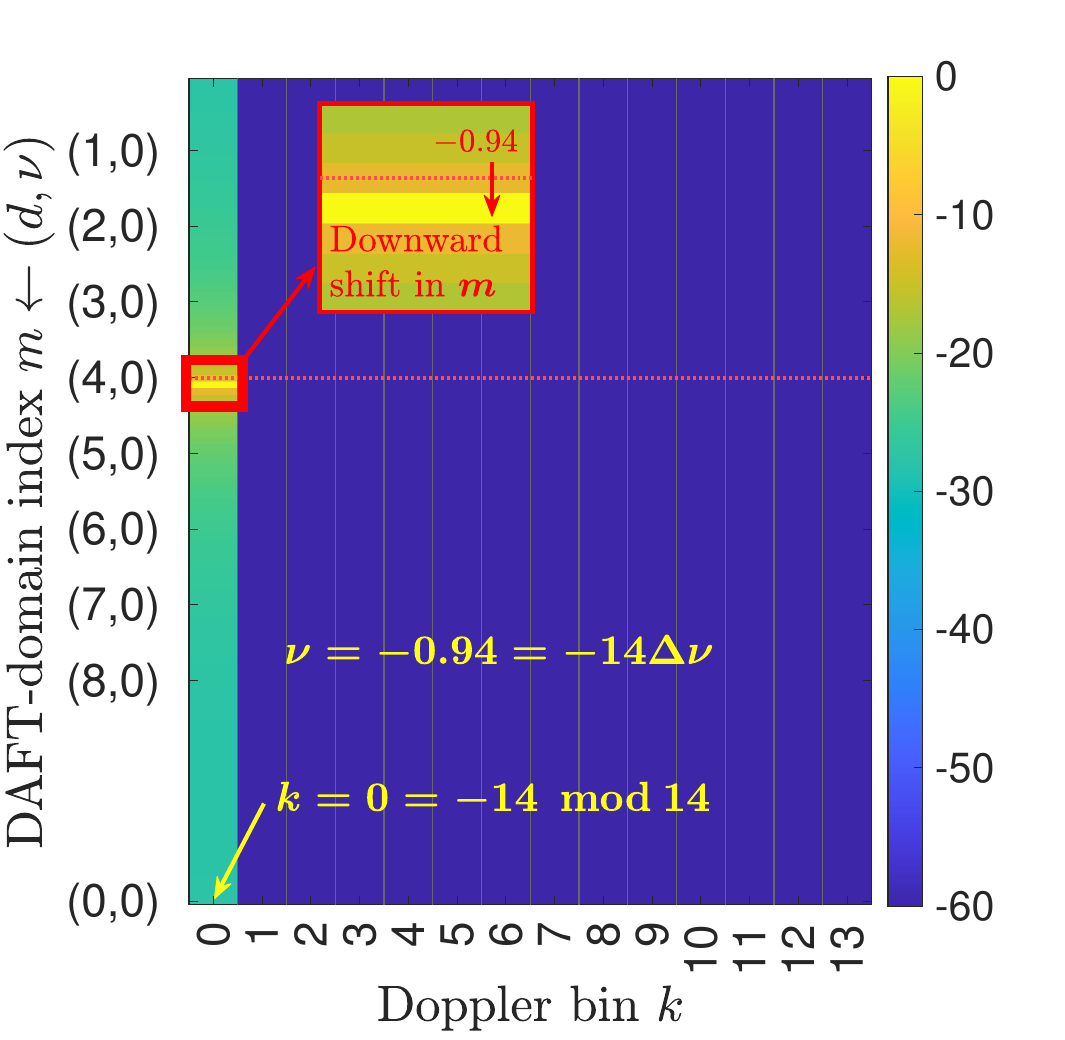}
			\label{fig:2Dmap-nu-094}
		}
		\caption{
			Visualization of $|\mat{W}|$ (in dB) for a noiseless scenario of one echo with $d_p=4$, $\nu_p=0.87$ (a) and $\nu_p=-0.94$ (b).
			The vertical axis highlights the DAFT-domain indices $m=-d_p\cdot c\mod M$ that correspond to delay-Doppler pairs $(d_p,\nu_p=0)$.
		}
		\label{fig:2Dmap-example}
		\vspace{-2.5em}
	\end{figure}

	Fig.~\ref{fig:2Dmap-example} shows two examples of 2D map $\mat{W}$ in a scenario with a single echo with delay $d=4$, and Doppler values  $\nu=0.87=13\Delta\nu$ (Fig.~\ref{fig:2Dmap-nu087}) and $\nu=-0.94=-14\Delta\nu$ (Fig.~\ref{fig:2Dmap-nu-094}).
	In these examples, $L=14$ and the Doppler values are intentionally chosen to be integer multiples of the Doppler bin resolution $\Delta\nu$~\eqref{eq:Doppler-Res}, so that the 2D map $\mat{W}$ exhibits non-zero components only for the Doppler bin indices $k=13 \mod L=13$ (Fig.~\ref{fig:2Dmap-nu087}), and $k=-14 \mod L=0$ (Fig.~\ref{fig:2Dmap-nu-094}).
	As shown in the previous examples, leakage occurs on the DAFT-domain indices since the Doppler values are not integers, resulting in shifted peak according the Doppler.
	
	\begin{figure}
		\centering
		\subfloat[$L=14$.]{
			\includegraphics[width=0.49\columnwidth]{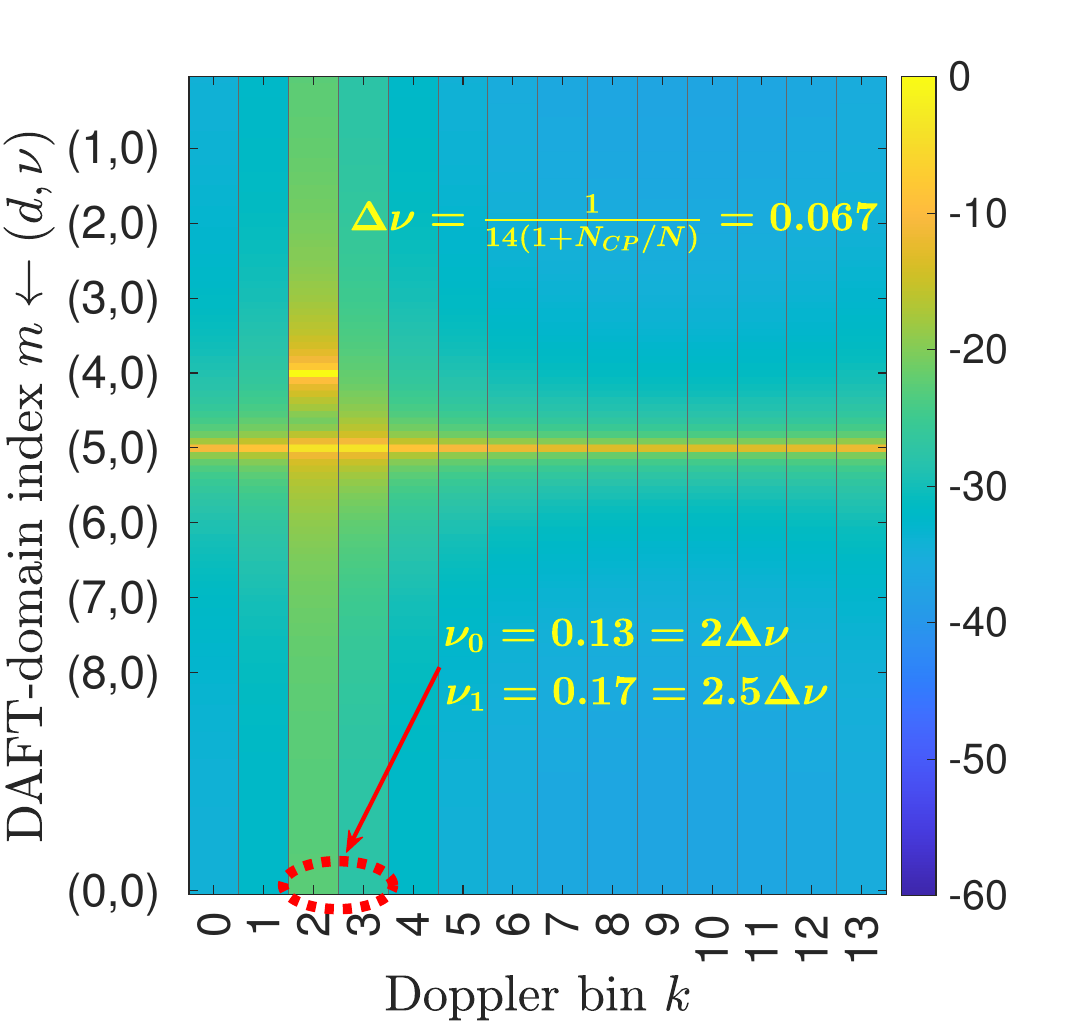}
			\label{fig:2Dmap-L}
		}
		\subfloat[$L=28$.]{
			\includegraphics[width=0.49\columnwidth]{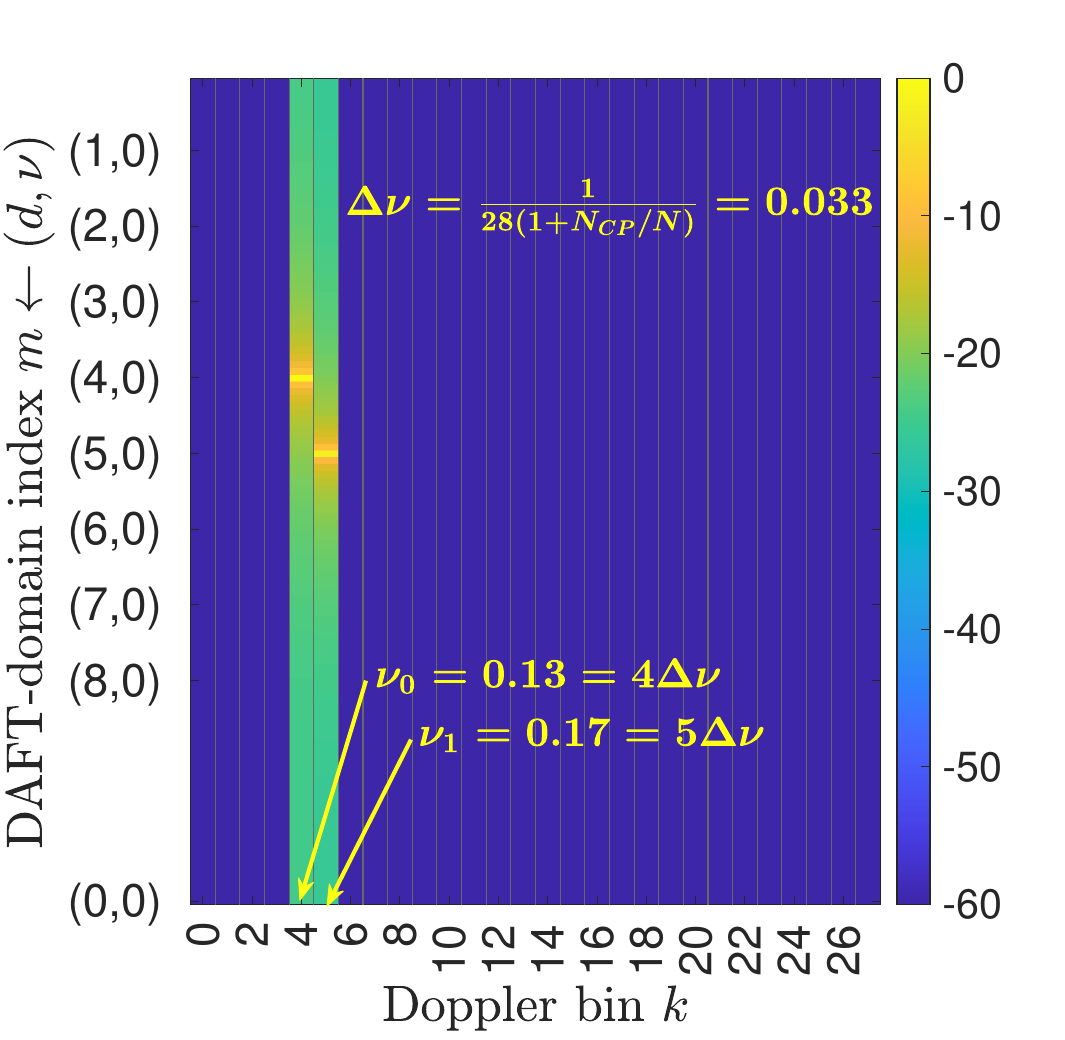}
			\label{fig:2Dmap-2L}
		}
		\caption{
			Visualization of $|\mat{W}|$ (in dB) for a noiseless scenario of two echoes with $d_p=[4,5]$ and $\nu_p=[0.13,0.17]$, for $L=14$ (a) and $L=28$ (b).
			The vertical axis highlights the DAFT-domain indices $m=-d_p\cdot c\mod M$ that correspond to delay-Doppler pairs $(d_p,\nu_p=0)$.
		}
		\label{fig:2Dmap-example-L}
	\end{figure}

	Fig.~\ref{fig:2Dmap-example-L} shows two examples of 2D map $\mat{W}$ in a scenario with two echoes with delays $d_p=[4,5]$ and Doppler values  $\nu_p=[0.13,0.17]$, for $L=14$ (Fig.~\ref{fig:2Dmap-L}) and $L=28$ (Fig.~\ref{fig:2Dmap-2L}) OFDM symbols.
	In Fig.~\ref{fig:2Dmap-L}, the echo $p=1$ manifests as a leakage over the Doppler bin indices since the Doppler value $\nu_1=0.17=2.5\Delta\nu$ is not an integer multiple of the Doppler resolution $\Delta\nu$.
	By increasing the number of OFDM symbols $L$, the Doppler resolution $\Delta\nu$~\eqref{eq:Doppler-Res} can be enhanced.
	For example, in Fig.~\ref{fig:2Dmap-2L}, the two echoes are separated on two distinct Doppler bin indices since both Doppler values are integer multiples of the resolution, i.e., $\nu_0,\nu_1=[4,5]\Delta\nu$.

\section{Results}
\label{sec:results}

	\begin{figure}
		\centering
		\subfloat[$\rho=1$.]{
			\includegraphics[width=0.49\columnwidth]{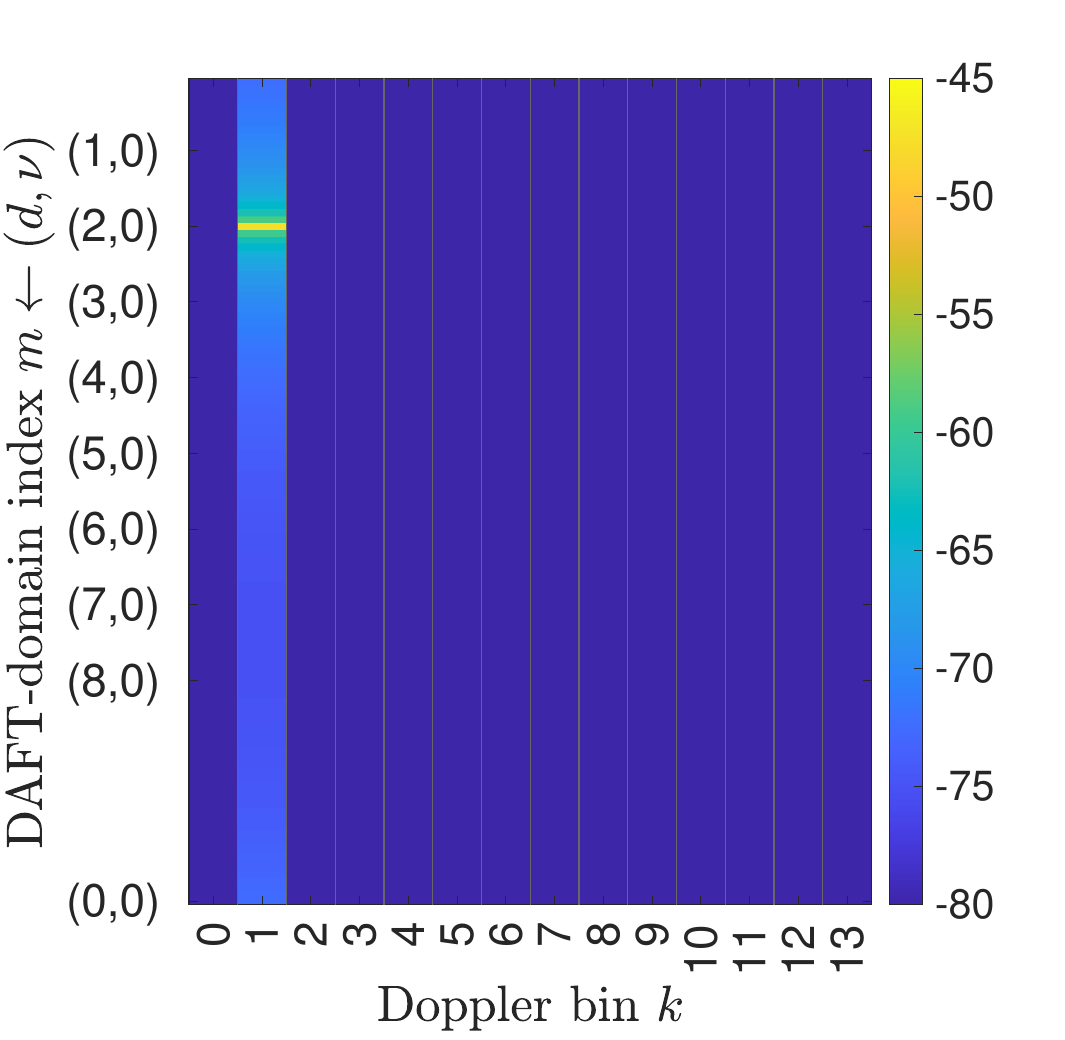}
			\label{fig:2Dmap-rho1}
		}
		\subfloat[$\rho=0.98$.]{
			\includegraphics[width=0.49\columnwidth]{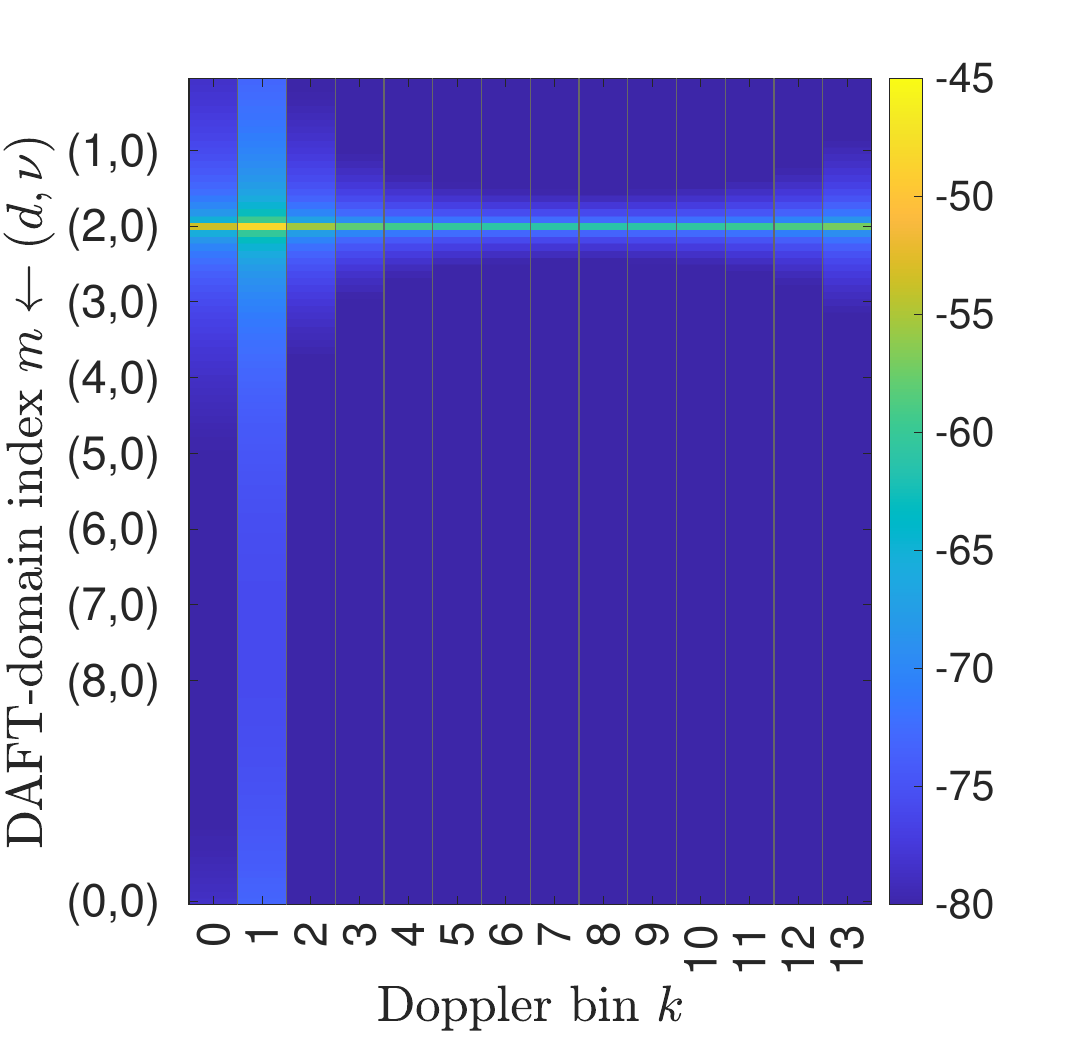}
			\label{fig:2Dmap-rho098}
		}
		\caption{
			2D maps $|\mat{W}|$ (dB scale):
			(a) $\rho=1$ (no RCS fluctuations);
			(b) $\rho=0.98$. 
		}	
		\label{fig:2Dmap-example-3}
	\end{figure}
	
	We employ a point scatterer model, where the channel gain coefficient $h_{p,l}$ incorporates the radar range equation to account for both the deterministic path loss and stochastic fluctuations in the received signal~\cite{RadarBook}.
	For the $p$-th target during the $l$-th OFDM symbol, the complex channel gain $h_{p,l}$ is modeled as
	\begin{equation}
		h_{p,l} = \sqrt{\frac{P_t G_t G_r c_0^2\; \sigma_{\text{RCS},p,l}}{(4\pi)^3 f_c^2 R_p^2}} e^{j\psi_{p,l}},
		\label{eq:h_pl}
	\end{equation}
	where $P_t$ is the transmitted power; 
	$G_t$ and $G_r$ are the antenna gains for the transmitter and receiver, respectively;
	$c_0$ is the speed of light;
	$f_c$ is the carrier frequency;
	$R_p$ is the range (distance) of the $p$-th target;
	$\sigma_{\text{RCS},p,l}$ is the radar cross section (RCS) magnitude and $\psi_{p,l}$ is the RCS phase~\cite{RadarBook}.
	The SNR is defined as $\gamma=\norm{\mat{H}\vect{x}}^2/\sigma_{\text{AWGN}}^2$.

	We consider a monostatic radar system observing a single drone target ($P=1$) with average RCS magnitude $\bar{\sigma}_{\text{RCS},p}=0.1$~m\textsuperscript{2}, which is typical of commercial UAVs~\cite{Tsai-2016}.
	The simulation parameters are: $P_t=1$~W, $G_t\cdot G_r=10$~dB, $f_c=7$~GHz, $\Delta f=30$~kHz, $N=M=120$, $\Ncp=8$. 
	The number of OFDM symbols is $L\in\{14,28\}$ depending on the experiment.
	The received echo is characterized by sample delay $d=2$ and normalized Doppler $\nu=1\cdot\Delta\nu=0.067$, corresponding to a target at range $R=83$ meters moving with relative velocity $v_{\text{rel}}=43$ m/s (155 km/h).
	The modeled velocity aligns with the United States Federal Aviation Administration's standard speed limit of 45 m/s for drones.

	The RCS is modeled using a first-order auto-regressive process to characterize temporal fluctuations in~\eqref{eq:h_pl} across successive OFDM symbols~\cite{Blair-2014, Meller-2019}. 
	Specifically, the RCS magnitude follows $\sigma_{\text{RCS},p,l+1}=\left|\rho \sigma_{\text{RCS},p,l} + (1-\rho) \bar{\sigma}_{\text{RCS},p} \cdot a\right|$, and the RCS phase evolves as $\psi_{p,l+1} =\rho \psi_{p,l} + (1-\rho) 2\pi\nu_p \cdot b$, where $a,b\sim\mathrm{Uniform}[-0.5,+0.5]$ are independent random variables.
	The parameter $\rho$ controls the fluctuation level within a TTI, where $\rho=1$ corresponds to a constant RCS (no fluctuations), while $0<\rho<1$ introduces time-varying fluctuations in the RCS.
	For instance, when $\rho=0.98$, the RCS phase exhibits a maximum variation of approximately $2\%$ of the Doppler shift per OFDM symbol. 
	This leads to a maximum cumulative phase variation up to $56\%$ of the Doppler shift across a TTI of $L=28$ symbols.

	Fig.~\ref{fig:2Dmap-example-3} presents 2D maps $|\mat{W}|$ for a noiseless scenario ($\sigma_{\text{AWGN}}^2=0$) with RCS fluctuation levels $\rho\in\{1,0.98\}$.
	The case with $\rho=1$, shown in Fig.~\ref{fig:2Dmap-rho1}, represents ideal conditions where no RCS fluctuations occur within a TTI, allowing for a clear peak identification in the 2D map.
	In contrast, Fig.~\ref{fig:2Dmap-rho098} shows the impact of RCS fluctuations ($\rho=0.98$) within a TTI, where the target's delay-Doppler estimation becomes challenging due to energy leakage across multiple Doppler bins over consecutive DAFT-domain indices.

	In the following performance results, we employ a delay-Doppler estimation algorithm that processes the 2D map $\mat{W}$.
	The algorithm consists in the identification of the indices $(\hat{m},\hat{k})$ of the peak in the 2D map, as this corresponds to the maximum likelihood estimator~\cite{Braun-MLradarOFDM-2010}. 
	The peak indices $(\hat{m},\hat{k})$ are then converted to the delay-Doppler estimates $(\hat{d},\hat{\nu})$ as described in Section~\ref{sec:Rx-processing}, such that $\hat{m}=-\hat{d}\cdot c \mod M$ and $\hat{\nu}=\tilde{k}\Delta\nu$, where $\hat{k} = \tilde{k} \mod L$.
	We assume that the estimated Doppler is $|\hat{\nu}|<0.5$ to avoid ambiguities.
	The performance results are computed over $10^5$ TTIs.
	
	\begin{figure}
		\includegraphics[width=\columnwidth]{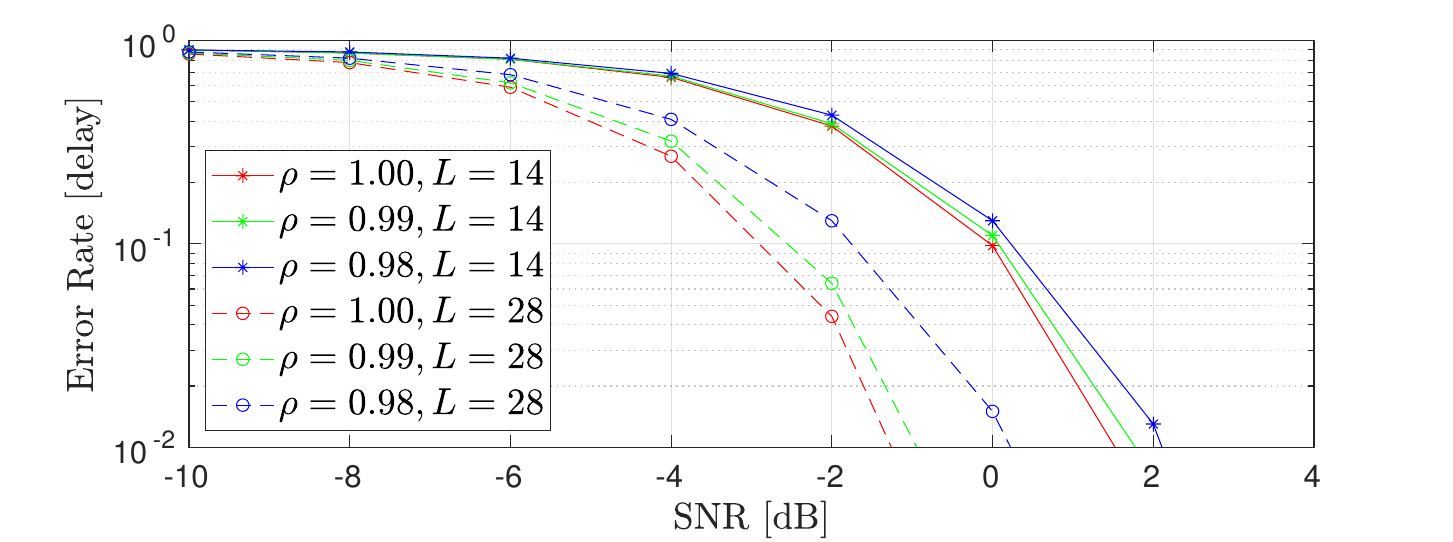}
		\caption{
			Error rate in delay estimation for $\rho\in\{1,0.99,0.98\}$ and $L\in\{14,28\}$.
		}
		\label{fig:perf-errRate}
		\vspace{-1em}
	\end{figure}
	
		\begin{figure}
		\includegraphics[width=\columnwidth]{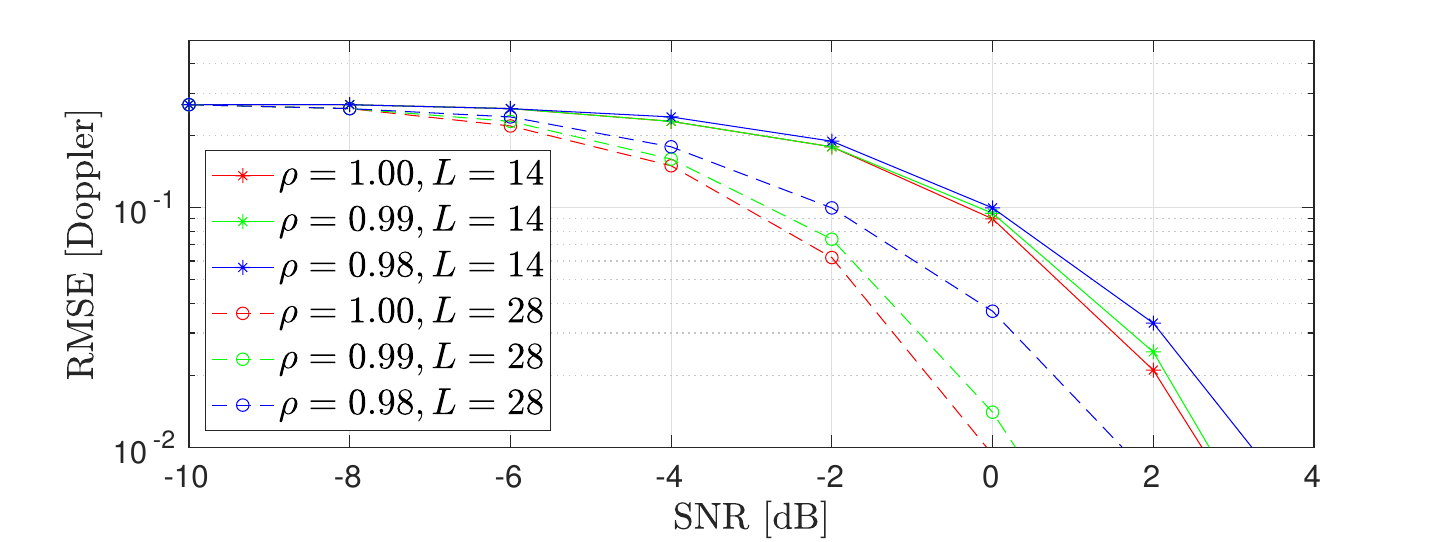}
		\caption{
			RMSE in Doppler estimation for $\rho\in\{1,0.99,0.98\}$ and $L\in\{14,28\}$.
		}
		\label{fig:perf-Doppler}
	\end{figure}

	Fig.~\ref{fig:perf-errRate} and Fig.~\ref{fig:perf-Doppler} show the error rate performance in delay estimation and the root mean squared error (RMSE) of Doppler estimation, respectively, under RCS fluctuation levels $\rho\in\{1,0.99,0.98\}$ with $L\in\{14,28\}$ OFDM symbols per TTI.
	Specifically, the error rate is computed as the ratio of error events to the total number of TTIs, where an error event occurs when the estimated delay is not equal to the ground truth.
	Both figures demonstrate the robustness of the DFT-p-FDMA-based ISAC framework, showing acceptable performance degradation despite signal fluctuations.

	For the ideal case without fluctuations ($\rho=1$), performance improves by approximately 3 dB when doubling $L$, as expected.
	With RCS fluctuations ($0<\rho<1$), increasing $L$ leads to progressive performance degradation relative to the fluctuation-free case. 
	At an error rate of 0.01 for delay estimation, the performance gap between $\rho=1$ and $\rho=\{0.99,0.98\}$ is $\{0.2,0.5\}$ dB for $L=14$, widening to $\{0.3,1.4\}$ dB for $L=28$. 
	Similarly, for Doppler estimation at an RMSE of 0.01, the corresponding gaps are $\{0.1,0.6\}$ dB ($L=14$) and $\{0.3,1.7\}$ dB ($L=28$).
	For $\rho=0.98$, the performance gap between $L=14$ and $L=28$ reduces to $\approx 2$ dB, demonstrating that while increasing the number of OFDM symbols generally improves estimation accuracy, the benefits diminish under stronger RCS fluctuations.
	These deteriorations result from accumulated RCS variations over longer TTIs and energy dispersion across multiple delay-Doppler components, as illustrated in Fig.~\ref{fig:2Dmap-rho098}.

\section{Conclusion}
\label{sec:conclusion}

	In this paper, we have proposed an ISAC framework based on the DFT-p-FDMA waveform.
	Our derived expression for the effective DFT-p-FDMA channel has established a direct relationship with the delay and Doppler characteristics of the received echoes.
	We have demonstrated that processing multiple received signals with a DFT operation enhances Doppler resolution. 
	Specifically, doubling the number of OFDM symbols halves the Doppler resolution, resulting in finer granularity and improved separation of echoes with distinct Doppler shifts.
	The examples and simulation results over various fluctuation and noise levels have validated the capabilities of the proposed framework for ISAC applications.

\appendices
\section{Derivation of the Effective DFT-p-FDMA Channel}
\label{sec:derivations}

	Consider~\eqref{eq:H-bar} to express the equivalent channel $\mat{\Hbar}$.
	Let $\mat{H}^\text{f}_l=\mat{F}_N \mat{H}_l \mat{F}_N^H$, where $\mat{H}_l$ is defined in~\eqref{eq:H}. 
	Then the $[k,i]$ element of $\mat{H}_l^\text{f}$ is 
	\begin{align}
		&\left( \mat{F}_N \mat{H}_l \mat{F}_N^H \right)[k,i] =\sum_{m,n} \mat{F}_N[k,m] \mat{H}_l[m,n] \mat{F}_N^H[n,i]\\
		&\propto\frac{1}{N}\sum_{m,n} e^{-j2\pi k \frac{m}{N}} e^{j2\pi \nu_p \frac{m}{N}} \delta_{m,n+d\mod N} \: e^{j2\pi i \frac{n}{N}}\\
		&\propto\frac{1}{N} e^{-j2\pi i \frac{d_p}{N}} \sum_{m}  e^{j2\pi \nu_p \frac{m}{N}(-k+\nu_p+i)} \\
		&\propto\frac{1}{N} e^{-j2\pi i \frac{d_p}{N}} D_N(\nu_p+i-k)\\
		&\mat{H}^\text{f}_l[k,i]=\sum_p \gamma(d_p,\nu_p,l) e^{-j2\pi i \frac{d_p}{N}} D_N(\nu_p+i-k),
	\end{align} 
	where $D_N(x)$ is defined in~\eqref{eq:kernel-D_N}.

	Assume $M=N$, which results in $\mat{S}=\mat{S}^T=\mat{I}$, and consider $\mat{P}$ such that $\pi_\mat{P}^{-1}=\pi_\mat{P}$, which corresponds to symmetry in $\mat{P}$, for simplicity and clarity in this derivation.
	Then, define $\mat{H}^\text{f-p}_l=\mat{P}^H \mat{S}^T \mat{H}^\text{f-p}_l  \mat{S}^T\mat{P}$, where the $[n,i]$ element of $\mat{H}^\text{f-p}_l$ is
	\begin{align}
		&\left( \mat{P}^H \mat{H}_l^\text{f} \mat{P} \right)[n,i] =\sum_{k,q} \mat{P}^H[n,k] \mat{H}^\text{f}_l[k,q] \mat{P}[q,i]\\
		&=\sum_{k,q} e^{-j\theta(k)} \delta_{n,\pi_{\mat{P}}(k)} \mat{H}^\text{f}_l[k,q] e^{j\theta(q)} \delta_{i,\pi_{\mat{P}}(q)}\\
		&=e^{-j\theta(\pi_\mat{P}(n))} \mat{H}^\text{f}_l[\pi_\mat{P}(n),\pi_\mat{P}(i)] e^{j\theta(\pi_\mat{P}(i))},
	\end{align}
	where $\mat{P}[a,b]$ is defined in~\eqref{eq:P-element} and $k=\pi_\mat{P}^{-1}(n)=\pi_\mat{P}(n)$.

	Finally, the effective channel for the $l$-th OFDM symbol~\eqref{eq:H-bar-closedForm} can be expressed as $\mat{\Hbar}_l=\mat{F}_M^H \mat{H}^\text{f-p}_l \mat{F}_M$, where the $[n,i]$ element of $\mat{\Hbar}_l$ is
	\begin{align}
		&\mat{\Hbar}_l[n,i] = \sum_{k,q} \mat{F}_M^H[n,k] \mat{H}^\text{f-p}_l[k,q] \mat{F}_M[q,i]\\
		&= \frac{1}{M} \sum_{k,q} e^{j2\pi \frac{nk-qi}{M}} e^{j\left[\theta(\pi_\mat{P}(q)) - \theta(\pi_\mat{P}(k))\right]} \mat{H}^\text{f}_l[\pi_\mat{P}(k),\pi_\mat{P}(q)].
	\end{align}

\balance
\bibliographystyle{IEEEtran}
\input{main.bbl}

\end{document}

%% file: commands.tex
\newcommand{\mat}[1]{\ensuremath{\mathbf{#1}}}
\newcommand{\vect}[1]{\ensuremath{\mathbf{#1}}}

\newcommand{\norm}[1]{||#1||}

\newcommand{\xtilde}{\widetilde{x}}
\newcommand{\ytilde}{\widetilde{y}}
\newcommand{\ybar}{\bar{y}}
\newcommand{\Ybar}{\bar{Y}}

\newcommand{\Hbar}{\bar{H}}

\newcommand{\Ncp}{{N_{\text{CP}}}}

\newcommand{\myBlue}{blue!80!black}
\newcommand{\myGreen}{green!60!black}
\newcommand{\myRed}{red!80!black}

\usepackage{tikz}

\usetikzlibrary{shapes,arrows,calc,positioning,fit,automata}
\usetikzlibrary{decorations.markings}
\usetikzlibrary{overlay-beamer-styles}
\usetikzlibrary{decorations.pathreplacing}

\tikzstyle{block}=[rectangle,draw,very thick,fill=white,align=center,font=\Large]
\tikzstyle{edge} = [draw,very thick,->,-triangle 45]
\tikzstyle{plateBlue} = [draw=\myBlue, shape=rectangle, rounded corners=0.5ex, ultra thick,
minimum width=3.1cm, text=\myBlue, text width=3.1cm, align=right, inner sep=10pt, inner ysep=10pt,append after command={node[font=\bf,text=\myBlue,above left= 3pt of \tikzlastnode.south east] {#1}}]

\tikzstyle{plateRed} = [draw=\myRed, shape=rectangle, rounded corners=0.5ex, ultra thick,
minimum width=3.1cm, text=\myRed, text width=3.1cm, align=right, inner sep=10pt, inner ysep=10pt,append after command={node[font=\bf\LARGE,text=\myRed,above= 3pt of \tikzlastnode.south] {#1}}]

\tikzstyle{plateGreen} = [draw=\myGreen, shape=rectangle, rounded corners=0.5ex, ultra thick,
minimum width=3.1cm, text=green!80!black, text width=3.1cm, align=right, inner sep=10pt, inner ysep=10pt,append after command={node[font=\bf\LARGE,text=\myGreen,above= 3pt of \tikzlastnode.south] {#1}}]

\tikzstyle{plate} = [draw, shape=rectangle, rounded corners=0.5ex, ultra thick,
minimum width=3.1cm,  text width=3.1cm, align=right, inner sep=10pt, inner ysep=10pt,append after command={node[font=\bf\Large,above left= 3pt of \tikzlastnode.south east] {#1}}]

\tikzstyle{plateSmall} = [draw, shape=rectangle, rounded corners=0.5ex, ultra thick,
minimum width=3.1cm,  text width=3.1cm, align=right, inner sep=10pt, inner ysep=10pt,append after command={node[font=\bf,above= 3pt of \tikzlastnode.south] {#1}}]

\tikzstyle{plateDashed} = [draw, dashed, shape=rectangle, rounded corners=0.5ex, ultra thick,
minimum width=3.1cm, text width=3.1cm, align=right, inner sep=10pt, inner ysep=10pt,append after command={node[font=\bf\LARGE,above= 3pt of \tikzlastnode.north] {#1}}]

%% file: Block_diagram_DFTpFDMA.tex
	\begin{tikzpicture}
		\def\dhor{2.8}
		\def\dver{0}
		
		\node[block, inner sep=0pt, draw=white] (input) at (0.25*\dhor,0) {};
		\node[block, inner sep=5pt, minimum height=70] (tx0) at (1*\dhor,0) {$M$-DFT\\$\mat{F}_M$};
		\node[block, inner sep=5pt, minimum height=70, draw=\myRed, line width=2] (tx0a) at (2.1*\dhor,0) {Phase rot.\\perm.\\$\mat{P}$};
		\node[block, inner sep=5pt, minimum height=70] (tx1) at (3.1*\dhor,0) {SC\\map\\$\mat{S}$};
		\node[block, inner sep=5pt, minimum height=70] (tx2) at (4*\dhor,0) {$N$-IDFT\\$\mat{F}_N^H$};
		\node[block, inner sep=5pt, minimum height=70] (tx3) at (5*\dhor,0) {Add\\CP};

		\node[block, inner sep=5pt, minimum height=50] (ch) at (6*\dhor,-1*\dver) {Channel\\$\mat{H}$};
		\node[block, circle, font=\bfseries\huge, inner sep=0pt] (add) at (6.75*\dhor,-2*\dver) {+};
		\node[block, draw=white, inner sep=0pt] (awgn) at (6.75*\dhor, 1.25) {AWGN};
		
		\node[block, inner sep=5pt, minimum height=70] (rx3) at (7.5*\dhor,-3*\dver) {Rem.\\CP};
		\node[block, inner sep=5pt, minimum height=70] (rx2) at (8.5*\dhor,-3*\dver) {$N$-DFT\\$\mat{F}_N$};
		\node[block, inner sep=5pt, minimum height=70] (rx1) at (9.5*\dhor,-3*\dver) {SC\\demap\\$\mat{S}^T$};
		\node[block, inner sep=5pt, minimum height=70, draw=\myRed, line width=2] (rx0a) at (10.75*\dhor,-3*\dver) {Phase de-rot.\\de-perm.\\$\mat{P}^H$};
		\node[block, inner sep=5pt, minimum height=70] (rx0) at (12*\dhor,-3*\dver) {$M$-IDFT\\$\mat{F}_M^H$};
		\node[block, inner sep=5pt, minimum height=70] (output) at (13*\dhor,-3*\dver) {Sensing};

		\path[edge] (ch) -- (add);
		\path[edge] (awgn) -- (add);
		
		\path[edge] (input) --node[block, inner sep=0, draw=white,above=0.2]{$\vect{u}_l$} node[block, inner sep=0, draw=white,below=0.2]{} (tx0);
		\path[edge] (tx0) -- (tx0a);
		\path[edge] (tx0a) --node[block, inner sep=0, draw=white,above=0.2]{$\vect{\xtilde}_l$} (tx1);
		\path[edge] (tx1) -- (tx2);
		\path[edge] (tx2) --node[block, inner sep=0, draw=white,above=0.2]{$\vect{x}_l$} (tx3);
		\path[edge] (tx3) -- (ch);
		
		\path[edge] (add) -- (rx3);
		\path[edge] (rx3) --node[block, inner sep=0, draw=white,above=0.2]{$\vect{y}_l$} (rx2);
		\path[edge] (rx2) -- (rx1);
		\path[edge] (rx1) --node[block, inner sep=0, draw=white,above=0.2]{$\vect{\ytilde}_l$} (rx0a);
		\path[edge] (rx0a) -- (rx0);
		\path[edge] (rx0) --node[block, inner sep=0, draw=white,above=0.2]{$\vect{\ybar}_l$} (output);
		
		\draw [decorate,decoration={brace,amplitude=15pt,mirror,raise=0ex}, line width=2, draw=\myBlue]
		(0.7*\dhor,-1.5) -- (12.3*\dhor,-1.5) node[midway,yshift=-3em,font=\LARGE, text=\myBlue]{$\mat{\Hbar}$};
		
	\end{tikzpicture}